\documentclass[fp,twocolumn]{jpsj3}
\usepackage{txfonts}
\usepackage{color}

\makeatletter
\providecommand{\ext@table}{lot}
\makeatother
\usepackage{etoolbox}
\usepackage[mathlines]{lineno}

\begin{document}

\title{
Ultrasonic Observation of Slowing Down of Multipole Fluctuations in Sr$_2$RuO$_4$
}

\author{
Ryosuke Kurihara$^1$
\thanks{
r.kurihara@rs.tus.ac.jp
or
iron.pnictide.man@gmail.com
},
Mitsuhiro Akatsu$^2$,
Shunsuke Yaoita$^3$,
Keisuke Mitsumoto$^4$,
Yuichi Nemoto$^3$,
Yoshiyuki Yoshida$^5$,
Hiroshi Yaguchi$^1$,
and
Terutaka Goto$^3$
}

\inst{
$^1$Department of Physics and Astronomy, Tokyo University of Science, Noda, Chiba 278-8510, Japan
\\
$^2$Department of Physics, Niigata University, Ikarashi, Niigata 950-2181, Japan
\\
$^3$Graduate School of Science and Technology, Niigata University, Ikarashi, Niigata 950-2181, Japan
\\
$^4$Liberal Arts and Sciences, Toyama Prefectural University, Imizu, Toyama 939-0398, Japan
\\
$^5$National Institute of Advanced Industrial Science and Technology (AIST), Tsukuba, Ibaraki 305-8568
}

\abst{
We performed ultrasonic measurements on the unconventional superconductor Sr$_2$RuO$_4$ to investigate the dynamical properties of the electronic states near its superconducting transition temperature, $T_\mathrm{c} = 1.4$ K.
We observed an increase in the in-plane transverse ultrasonic attenuation coefficient as the temperature approached $T_\mathrm{c}$.
The ultrasonic attenuation exhibited a Landau-Khalatnikov-type ultrasonic frequency dependence with a typical relaxation time of approximately $10^{-10}$ s.
Under an applied magnetic field of 10 T, the superconducting transition was suppressed.
However, the ultrasonic attenuation coefficient exhibited an increase down to low temperatures, indicating the slowing down of fluctuations associated with multipole degrees of freedom. 
Based on group-theoretical considerations, we propose that the electric hexadecapole plays a crucial role in the slowing down.
Furthermore, we discuss the relationship between multi-component superconducting order parameters and multipole degrees of freedom.
}
\maketitle

\section{
Introduction
}
\label{sect_intro}

Since its discovery, superconductivity has been the subject of intensive research in fundamental physics.
Understanding the mechanism of superconductivity can lead to the discovery of new types of superconductors with high transition temperatures
\cite{Ashcroft_PRL9, Drozdov_Nature525, Anderson_Science235, Steglich_PRL43, Kamihara_JACS130}.
Superconductivity also plays an essential role in industrial applications, including superconducting magnets
\cite{Maeda_IEEE24},
quantum devices
\cite{Gao_PRXQ2}, 
and motors for hydrogen-powered vehicles operating at liquid-hydrogen temperature
\cite{Toyota_Superconductor}.  
From a fundamental point of view, investigating the symmetry of the superconducting order parameter and the pairing mechanism is crucial.
To this end, physical property measurements have been performed to investigate the normal and superconducting states
\cite{Ishida_JPSJ62, Mignod_PhysC185, Shishido_JPSJ74, Yoshizawa_JPSJ81}.
These results can provide insight into the symmetry of the superconducting order parameter and the superconducting mechanism
\cite{Ishida_Nature396, Ghosh_NatPhys17, Benhabib_NatPhys17}.

Sr$_2$RuO$_4$ is one of the superconductors for which the symmetry of the superconducting order parameter has been extensively discussed.
The discovery of superconductivity in Sr$_2$RuO$_4$ by Maeno \textit{et al.}
\cite{Maeno_Nature372},
has received much attention because its layered perovskite structure with the space group of $I4/mmm$ ($D_{4h}^{17}$, No. 139) is the same as that of copper-oxide superconductors.
Sr$_2$RuO$_4$ exhibits a superconducting transition at $T_\mathrm{c} = 1.4$ K
\cite{Mackenzie_PRL80},
and Ru-$4d$ electrons contribute to its multiband electronic structure by forming Fermi surfaces (FSs)
\cite{Bergemann_PRL84, Oguchi_PRB51, Tamai_PRX9}.
Many experimental and theoretical studies have been conducted to shed light on the superconducting mechanism in this non-cuprate layered perovskite system
\cite{Maeno_JPSJ93, Mackenzie_npj2}.
However, the superconducting mechanism of Sr$_2$RuO$_4$ remains under active debate 
\cite{Ishida_Nature396, Pustogow_Nature574, Ishida_JPSJ89, Maeno_NatPhys20, Kittaka_JPSJ95, Matsuki_PRL136}.

Recently, the relationship between the elasticity and superconducting properties in Sr$_2$RuO$_4$ attracted considerable attention.
Previous studies have shown that the elastic constants exhibit anomalies around $T_\mathrm{c}$
\cite{Matsui_JPSJ67, Okuda_JPSJ71, Lupien_PRL86, Lupien_DrThesis, Matsui_PRB63, Ghosh_NatPhys17, Benhabib_NatPhys17}.
Based on Landau phenomenological theory, we can describe the free energy of superconductivity, including the coupling of the superconducting order parameter and strain
\cite{Luthi_Text}.
Considering the irreducible representation (irrep) $\Gamma$ of the elastic constant $C_\Gamma$, which exhibits a discontinuous anomaly at $T_\mathrm{c}$, we can deduce the possible symmetries of the order parameters $\Gamma'$ and $\Gamma''$ whose direct product $\Gamma' \otimes \Gamma'' $ contains $\Gamma$.
Ultrasonic measurements based on this phenomenological description have proposed several types of multi-component superconducting order parameters
\cite{Ghosh_NatPhys17, Benhabib_NatPhys17}.
The dependence of the superconducting transition temperature on external strain has also played a key role in determining the superconducting order parameters
\cite{Steppke_Science355, Hicks_Science344, Li_Nature607, Mattoni_NatCommun}. 

In addition to the elastic constants, the ultrasonic attenuation coefficient, which corresponds to the imaginary part of the complex elastic constant, has played a key role in investigating superconductivity
\cite{BCS_PR108, Tinkham_Text, Bliss_PR177, Phillips_PRSA309, Bishop_PRL53, Batlogg_PRL55, Bishop_PRB35}.
Recent ultrasonic studies have also revealed significant ultrasonic attenuation around $T_\mathrm{c}$ originating from the critical slowing down of the electric multipole degree of freedom in the iron-based superconductor Ba(Fe,Co)$_2$As$_2$ with a multiband electronic structure
\cite{Kurihara_JPSJ86, Sato_JPSJ92}.
Therefore, it is worth investigating the possibility of the slowing down of multipole fluctuations around $T_\mathrm{c}$ in Sr$_2$RuO$_4$.
The ultrasonic attenuation coefficients in the superconducting phase of Sr$_2$RuO$_4$ have been intensively investigated
\cite{Lupien_DrThesis, Lupien_PRL86, Matsui_PRB63}.
However, a detailed investigation of ultrasonic attenuation in the normal state is still lacking.
Furthermore, this could confirm multi-component superconductivity within the framework of group theory.

This paper is organized as follows:
In Sec. \ref{sect_exp}, details of ultrasonic measurements are shown.
In Sec. \ref{Results}, we present the ultrasonic attenuation coefficient and the elastic constant of Sr$_2$RuO$_4$.
We demonstrate that anomalous ultrasonic attenuation occurs at low temperatures.
In Sec. \ref{Discussion}, we demonstrate that the ultrasonic attenuation originates from the slowing down of fluctuations of electronic degrees of freedom.
Its dynamical properties, candidate degrees of freedom, described as the electric hexadecapole, and possible contributions to the superconducting mechanism of Sr$_2$RuO$_4$ are also discussed in the framework of group theory. 
Our conclusions on the ultrasonic measurements and the slowing down in Sr$_2$RuO$_4$ are presented in Sec. \ref{conclusion}.

\section{
Experimental setup
}
\label{sect_exp}

Single crystals of Sr$_2$RuO$_4$ were grown by the floating-zone method using an ellipsoidal image furnace.
An X-ray back-scattering method with a Laue camera was used to characterize the crystallographic orientations.
The mass density, $\rho = 5.96$ g/cm$^3$
\cite{Walz_ActaCrystC49},
and the ultrasonic velocity, $v$, were used to estimate the elastic constant, $C = \rho v^2$.
X-cut and 36$^\circ$ Y-cut LiNbO$_3$ piezoelectric transducers were used to generate the transverse and longitudinal ultrasonic waves, respectively.
Transducers with a thickness of 100 $\mu$m (40 $\mu$m) and fundamental frequencies of approximately 18 (40) MHz were used for the transverse waves, while those with a thickness of 100 $\mu$m and a fundamental frequency of 33 MHz were used for the longitudinal waves.
Transducers were attached using silicone rubber (Shin-Etsu Silicone, KE-4897T).
The propagation direction $\boldsymbol{q}$ and the polarization direction $\boldsymbol{\xi}$ corresponding to each ultrasonic attenuation coefficient and elastic constant are listed in Table \ref{table_q_and_xi}. 

Both numerical vector-type phase-detection and analog-type frequency-modulation techniques were employed to determine the sound velocity and the ultrasonic attenuation coefficient as described in the Supplemental Material
\cite{Kurihara_SM}.
A $^3$He cryostat equipped with a superconducting magnet providing magnetic fields of up to 10 T (Oxford Instruments, Heliox TL) was used for measurements down to 300 mK.

\begin{table}[b]
\caption{
Propagation direction $\boldsymbol{q}$ and polarization direction $\boldsymbol{\xi}$ of the ultrasonic waves corresponding to each elastic constant.
The abbreviations L and T in the column labeled "Type" indicate longitudinal and transverse ultrasonic waves, respectively.  
}
\label{table_q_and_xi}
\begin{center}
\begin{tabular}{cccc}
\hline
$\boldsymbol{q}$
    &$\boldsymbol{\xi}$
        &Elastic constant
            &Type
\\
\hline
$[110]$
    &$[1\overline{1}0]$
        &$C_\mathrm{T} = \left( C_{11} - C_{12} \right)/2$
            &T
\\
$[100]$
    &$[100]$
        &$C_{11}$
            &L
\\
$[100]$
    &$[001]$
        &$C_{44}$
            &T
\\
$[100]$
    &$[010]$
        &$C_{66}$
            &T
\\  \hline
\end{tabular}
\end{center}
\end{table}

\section{
Results
}
\label{Results}

\begin{figure}[t]
\begin{center}
\includegraphics[clip, width=0.42\textwidth]{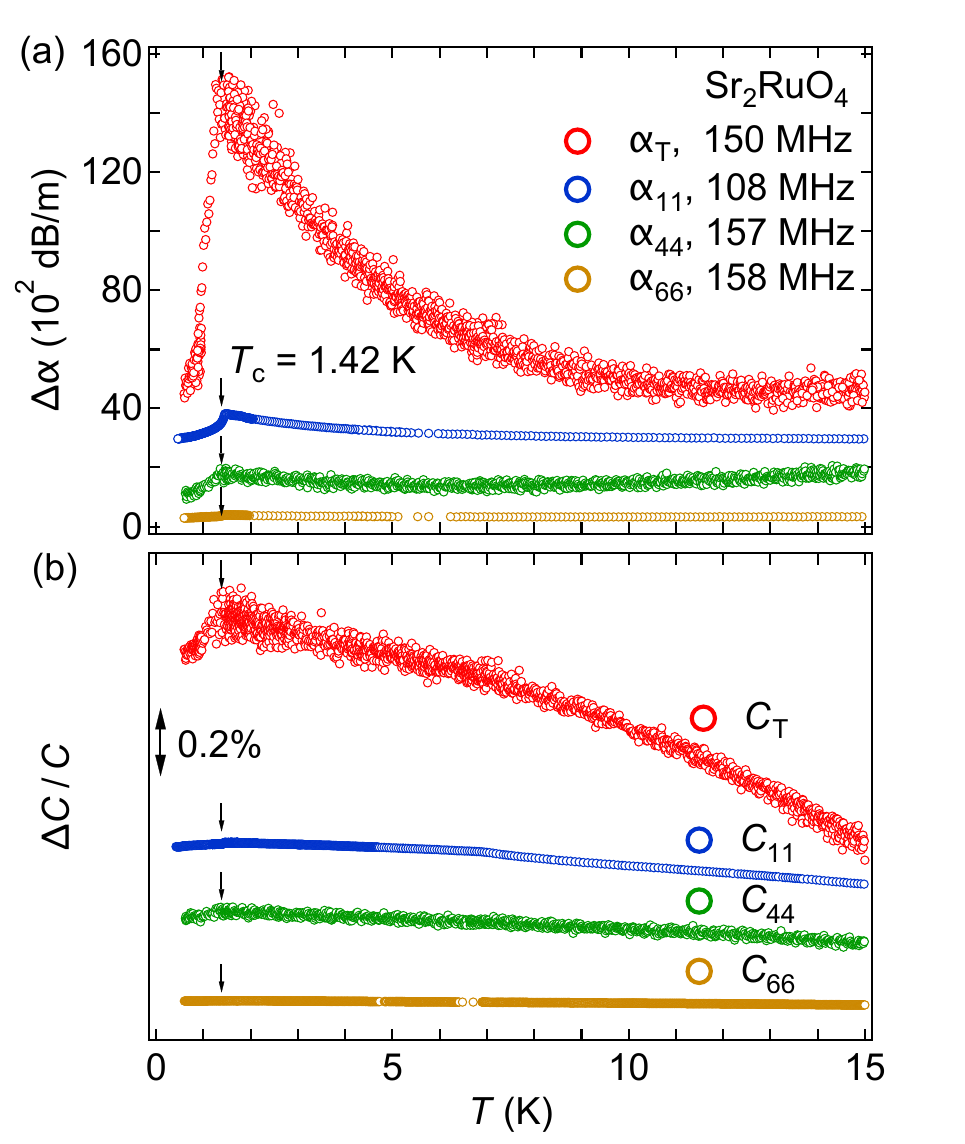}
\end{center}
\caption{
(Color online)
(a) Temperature dependence of the change in the ultrasonic attenuation coefficients $\Delta \alpha_\mathrm{T}$, $\Delta \alpha_\mathrm{11}$, $\Delta \alpha_\mathrm{44}$, and $\Delta \alpha_\mathrm{66}$, and (b) the relative change in the elastic constants $\Delta C_\mathrm{T} / C_\mathrm{T}$, $\Delta C_\mathrm{11} / C_\mathrm{11}$, $\Delta C_\mathrm{44} / C_\mathrm{44}$, and $\Delta C_\mathrm{66} / C_\mathrm{66}$.
The ultrasonic frequency for each $\Delta \alpha$ and $\Delta C / C $ is indicated.
Vertical arrows indicate the superconducting transition temperature, $T_\mathrm{c} = 1.42$ K. 
The data sets are vertically offset for clarity.
}
\label{Fig_Attenuation_all}
\end{figure}

Figure \ref{Fig_Attenuation_all}(a) shows the temperature dependence of the change in the ultrasonic attenuation coefficients of Sr$_2$RuO$_4$.
We observed a significant increase in $\Delta \alpha_\mathrm{T}$ from approximately 4000 dB/m at 15 K to about 16000 dB/m near $T_\mathrm{c} = 1.42$ K.
This attenuation coefficient corresponds to the elastic constant $C_\mathrm{T} = \left( C_{11} - C_{12} \right)/2$, associated with in-plane transverse ultrasonic waves with propagation direction $\boldsymbol{q} // [110]$ and polarization direction $\boldsymbol{\xi} // [1\overline{1}0]$.
No significant increase was observed in $\Delta \alpha_{11}$, $\Delta \alpha_{44}$, or $\Delta \alpha_{66}$.
On the other hand, we were unable to measure $\alpha_{33}$ associated with the inter-plane longitudinal waves.
This may be due to the layered crystal structure of Sr$_2$RuO$_4$.

In addition to the ultrasonic attenuation coefficients, we measured the temperature dependence of the elastic constants in Sr$_2$RuO$_4$, as shown in Fig. \ref{Fig_Attenuation_all}(b).
The relative change in the elastic constant, denoted as $\Delta C / C = \left[ C \left( T \right) - C \left( T_\mathrm{c} \right) \right] /  C \left( T_\mathrm{c} \right)$, shows a monotonic hardening below 15 K down to $T_\mathrm{c} = 1.42$ K, followed by a softening below $T_\mathrm{c}$.
This behavior is consistent with previous studies
\cite{Matsui_JPSJ67, Okuda_JPSJ71, Lupien_PRL86, Lupien_DrThesis, Matsui_PRB63, Ghosh_NatPhys17, Benhabib_NatPhys17}.

\begin{figure*}[t]
\begin{center}
\includegraphics[clip, width=0.8\textwidth]{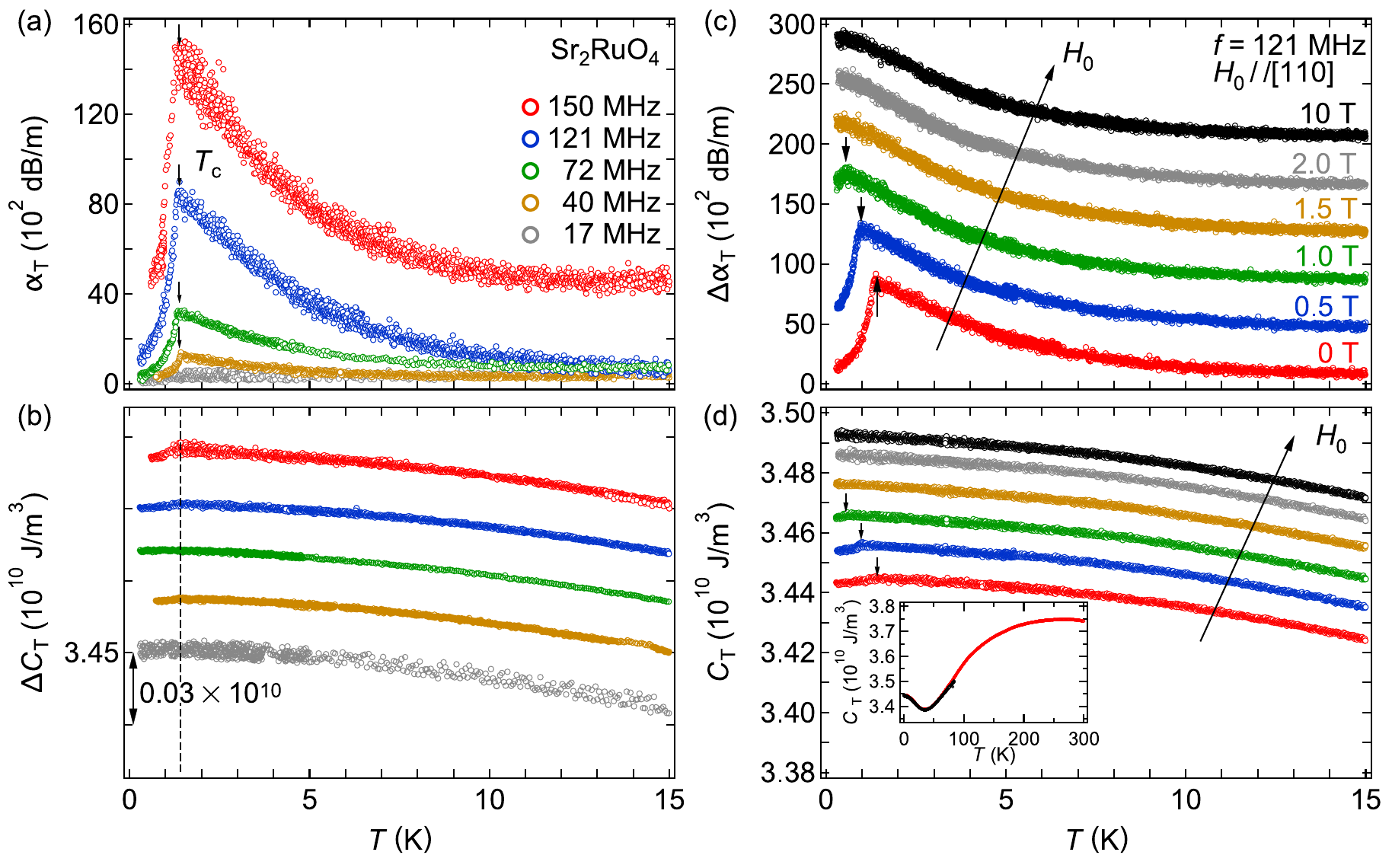}
\end{center}
\caption{
(color online)
Temperature dependence of (a) the ultrasonic attenuation coefficient $\alpha_\mathrm{T}$ and (b) the change in the elastic constant $C_\mathrm{T} = \left( C_{11} - C_{12} \right)/2$ of Sr$_2$RuO$_4$ measured at several ultrasonic frequencies.
Panels (c) and (d) show the temperature dependence of the change in the ultrasonic attenuation coefficient $\Delta \alpha_\mathrm{T}$ and the elastic constant $C_\mathrm{T}$, respectively, measured at an ultrasonic frequency of 121 MHz under a magnetic field applied along the [110] direction.
Vertical dashed lines and downward arrows in panel (b) indicate the superconducting transition temperature, $T_\mathrm{c} = 1.42$ K.
The $\Delta C_\mathrm{T}$ and $\Delta \alpha_\mathrm{T}$ data sets are vertically offset for clarity.
Each $C_\mathrm{T}$ data set is additionally shifted vertically by $1 \times 10^8$ J/m$^3$ for clarity.
The inset in panel (d) shows $C_\mathrm{T}$ at 0 and 10 T up to 300 K.
}
\label{Fig_CandAttenuations}
\end{figure*}

To further understand the ultrasonic attenuation, we measured the frequency dependence of $\alpha_\mathrm{T}$, as shown in Fig. \ref{Fig_CandAttenuations}(a).
The ultrasonic attenuation increases with increasing frequency, indicating a dynamic response of the electronic degrees of freedom. 
In contrast to $\alpha_\mathrm{T}$, $C_\mathrm{T}$ is nearly independent of frequency, except for the elastic anomaly at $T_\mathrm{c}$, as shown in Fig. \ref{Fig_CandAttenuations}(b).
This behavior may reflect the enhanced sensitivity of the sound velocity at high frequencies.

To clarify whether the ultrasonic attenuation originates from superconductivity in Sr$_2$RuO$_4$, we measured $\alpha_\mathrm{T}$ and $C_\mathrm{T}$ under a magnetic field, as shown in Figs. \ref{Fig_CandAttenuations}(c) and \ref{Fig_CandAttenuations}(d).
As the magnetic field increases, the decrease in $\Delta \alpha_\mathrm{T}$ associated with the superconducting transition is gradually suppressed.
Previous studies have reported similar results
\cite{Lupien_PRL86}.
Above the upper critical field $H_\mathrm{c2} = 1.5$ T
\cite{Akima_JPSJ68}, 
$\Delta \alpha_\mathrm{T}$ increases monotonically down to low temperatures.
$C_\mathrm{T}$ under a magnetic field exhibits similar behavior.
These results suggest that the increase in $\alpha_\mathrm{T}$ around $T_\mathrm{c}$ is not caused by the superconducting transition.
In other words, the increase in $\Delta \alpha_\mathrm{T}$ may not be fully explained by superconductivity alone, suggesting the involvement of additional electronic degrees of freedom.
However, since the superconducting transition occurs first, these degrees of freedom do not develop an independent ordered state, and the attenuation coefficient decreases below $T_\mathrm{c}$ due to superconductivity.
This interpretation is supported by simultaneous measurements of AC magnetic susceptibility using the mutual-induction method and elasticity using the ultrasonic pulse-echo method (see Supplemental Material \cite{Kurihara_SM}).
We note that the decrease in $\alpha_\mathrm{T}$ below $T_\mathrm{c}$ is consistent with the non-BCS-like temperature dependence previously reported
\cite{Lupien_PRL86, Matsui_PRB63}.
We discuss the possible origin of these degrees of freedom responsible for the ultrasonic attenuation in the following section.

We also measured $C_\mathrm{T}$ under a magnetic field of 10 T up to 80 K, as shown in the inset of Fig. \ref{Fig_CandAttenuations}(d).
$C_\mathrm{T}$ is robust against the magnetic field, retaining the softening, the minimum temperature, and the subsequent hardening down to $T_\mathrm{c}$.
This result suggests that the magnetic degrees of freedom do not contribute to the elastic properties.

Although the ultrasonic attenuation coefficient $\alpha_\mathrm{T}$ exhibits a pronounced increase, the elastic constant $C_\mathrm{T}$ does not show a corresponding softening in the same temperature range.
To understand the origin of this anomaly, we focus on the relaxation time of the electronic degrees of freedom.
In the presence of relaxational dynamics, the ultrasonic attenuation is expected to follow a Debye-type frequency dependence.
Based on the Landau-Khalatnikov theory
\cite{Luthi_Text},
we can describe the angular-frequency dependence of the ultrasonic attenuation coefficient as follows:
\begin{equation}
\label{Debye}
\alpha_\mathrm{T} \left( \omega \right)
= \frac{C_\mathrm{T}^\infty - C_\mathrm{T}^0 }{2 \rho \left( v_\mathrm{T}^\infty \right)^3}
\frac{\omega^2 \tau}{1 + \omega^2 \tau^2}
\xrightarrow[\omega \tau \ll 1]{}
\frac{C_\mathrm{T}^\infty - C_\mathrm{T}^0 }{2 \rho \left( v_\mathrm{T}^\infty \right)^3} \omega^2 \tau
.
\end{equation}
Here, $C_\mathrm{T}^\infty$ ($C_\mathrm{T}^0$) is the high-frequency (low-frequency) limit of the elastic constant $C_\mathrm{T}$, $v_\mathrm{T}^\infty = \sqrt{ C_\mathrm{T}^\infty / \rho }$ is the high-frequency limit of the sound velocity, $\omega$ is the angular frequency of ultrasonic waves, and $\tau$ is the relaxation time.
The relaxation time is assumed to be independent of the angular frequency.
We assume that $C_\mathrm{T}^0$ corresponds to the experimental result of $C_\mathrm{T}$ in Fig. \ref{Fig_CandAttenuations} and that $C_\mathrm{T}^\infty$ is described by the temperature-independent background component in $C_\mathrm{T}$.
Because $C_\mathrm{T}$ shows a softening below 270 K down to 35 K as shown in the inset of Fig. \ref{Fig_CandAttenuations}(d),
we regard the value at the onset of the softening as the background elastic constant.
Thus, we assume $C_\mathrm{T}^\infty = C_\mathrm{T} \left( T = 270 \ \mathrm{K} \right) = 3.75 \times 10^{10} \ \mathrm{J/m}^3$.

\begin{figure}[t]
\begin{center}
\includegraphics[clip, width=0.5\textwidth]{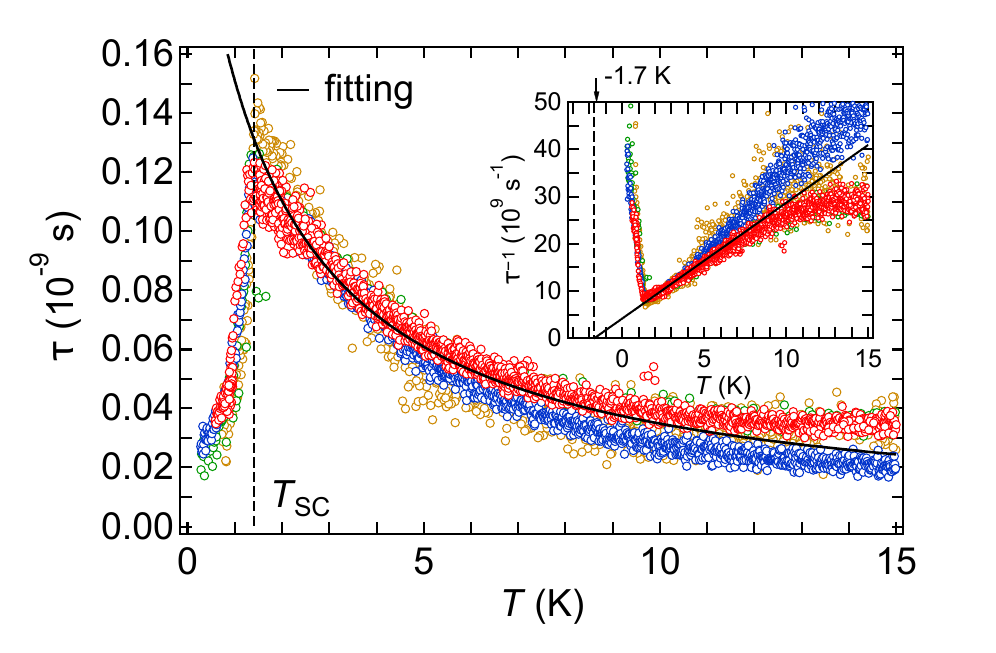}
\end{center}
\caption{
(color online)
Temperature dependence of the relaxation time $\tau$ calculated from the ultrasonic attenuation coefficient $\alpha_\mathrm{T}$ shown in Fig. \ref{Fig_CandAttenuations}(a), except for the 17 MHz data.
The marker colors indicate the frequency of $\alpha_\mathrm{T}$, and correspond to those used in Fig. \ref{Fig_CandAttenuations}(a).
The solid black line represents fits of $\tau$ using Eq.~(\ref{tau_MFT}).
The inset exhibits the temperature dependence of the relaxation rate $\tau^{-1}$.
}
\label{Fig_tau}
\end{figure}

Based on Eq.~(\ref{Debye}), we calculated the relaxation time $\tau$ for $\alpha_\mathrm{T}$ of Sr$_2$RuO$_4$ in Fig. \ref{Fig_tau}.
$\tau$ shows an increase from about $0.025 \times 10^{-9}$ s at 15 K up to about $0.14 \times 10^{-9}$ s at $T_\mathrm{c}$.
We note that the condition $\omega \tau \ll 1$ is satisfied for the ultrasonic frequencies below 150 MHz, and that $\tau$ exhibits frequency-independent behavior.
These results indicate that the ultrasonic attenuation is described by the Landau-Khalatnikov theory. 
We note that the $\omega^2$-dependence of $\alpha_\mathrm{T}$ is consistent with previous ultrasonic measurements 
\cite{Lupien_PRL86}.

As shown above, we observed the anomalous ultrasonic attenuation around the superconducting transition.
Furthermore, we revealed that the temperature dependence of the ultrasonic attenuation shows an increase at low temperatures under magnetic fields. 
In the following, we discuss the origin of this ultrasonic attenuation and its dynamic properties.

\section{
Analysis and Discussion
}
\label{Discussion}

To analyze the temperature dependence of $\tau$ in the normal state, we employ the following phenomenological description for the relaxation time
\cite{Nishimori_Text, Kurihara_JPSJ86}:
\begin{equation}
\label{tau_MFT}
\tau
= \tau_0 \varepsilon^{-z {\nu}}
= \tau_0 \left| \frac{ T - T_\mathrm{c}^0 } { T_\mathrm{c}^0 }  \right|^{-z \nu}
.
\end{equation}
Here, $\tau_0$ is a constant, $\varepsilon = \left| \left( T - T_\mathrm{c}^0 \right) / T_\mathrm{c}^0 \right|$ is the reduced temperature, $T_\mathrm{c}^0$ is the critical temperature, ${\nu}$ is the critical exponent for the correlation length, and $z$ is the dynamical critical exponent.

Equation (\ref{tau_MFT}) reproduces well the temperature dependence of the relaxation time above $T_\mathrm{c}$, as shown in Fig. \ref{Fig_tau}.
The obtained parameters are listed in Table \ref{table_parameters}.
The product of the critical exponents, $z {\nu} = 1$, is consistent with the mean-field prediction for the Ising universality class.
We note that $T_\mathrm{c}^0$ in the normal-state fit
takes a negative value of $-1.7$ K, suggesting that a ferro-type ordering of the electronic degrees of freedom does not occur.
This behavior contrasts with that observed in Ba(Fe,Co)$_2$As$_2$, where $T_\mathrm{c}^0$ was found to be nearly identical to the superconducting transition temperature, indicating the development of ferro-type hexadecapole ordering
\cite{Kurihara_JPSJ86}.

\begin{table}[b]
\caption{
Parameters obtained from fitting $\tau$ to Eq.~(\ref{tau_MFT}) separately in the normal (N) and superconducting (S) phases of Sr$_2$RuO$_4$ and iron-based superconductor Ba(Fe,Co)$_2$As$_2$ \cite{Kurihara_JPSJ86}.
}
\label{table_parameters}
\begin{center}
\begin{tabular}{ccccc}
\hline
Phase
    &$\tau_0$ (sec)
        &$\tau_0^{-1}$ (Hz)
            &$z \nu$
                &$T_\mathrm{c}^0$ (K)
\\
\hline
Sr$_2$RuO$_4$ (N)
    &$2.4 \times 10^{-10}$   
        &$0.42 \times 10^{10}$    
            &$1$
                &$-1.7$
\\
Ba(Fe,Co)$_2$As$_2$ (N)
    &$0.069 \times 10^{-10}$   
        &$14 \times 10^{10}$    
            &$1$
                &$23.0$
\\
Ba(Fe,Co)$_2$As$_2$ (S)
    &$0.067 \times 10^{-10}$   
        &$15 \times 10^{10}$     
            &$1/3$
                &$23.5$
\\  \hline
\end{tabular}
\end{center}
\end{table}

In addition to the critical exponents and critical temperatures, we can discuss the dynamical properties of the electronic degrees of freedom. 
As shown in the inset of Fig. \ref{Fig_tau}, the relaxation rate $\tau^{-1}$ decreases toward $T_\mathrm{c}$, indicating a slowing down of the fluctuations of the electronic degrees of freedom.
Due to this slowing down, $\tau^{-1}$ reaches 7.1 GHz, and the corresponding energy scale is estimated to be $h \tau^{-1} \sim 0.34$ K $\sim$ $29$ $\mathrm{\mu}$eV, where $h$ is the Planck constant.
Therefore, the electronic degrees of freedom with low-energy dynamics that form an Ising-type Hamiltonian described by mean-field theory can play a crucial role at low temperatures.

To further investigate the electronic degrees of freedom, we consider the deformation of a crystal lattice, denoted as $\boldsymbol{\xi} \left( \boldsymbol{r}, t \right)$, which is introduced by the transverse ultrasonic waves with $\boldsymbol{q} // [110]$ and $\boldsymbol{\xi} // [1\overline{1}0]$ for $\alpha_\mathrm{T}$ and $C_\mathrm{T}$
\cite{Luthi_Text}.
Here, $\boldsymbol{r} = \left(x, y, z \right)$ is the position vector in the elastic medium.
Assuming that the deformation is described by the plane wave formula, we can express $\boldsymbol{\xi} \left( \boldsymbol{r}, t \right)$ as follows:  
\begin{equation}
\label{xi_vector}
\boldsymbol{\xi} \left( \boldsymbol{r}, t \right)
= \boldsymbol{\xi}_0 \exp\left[ i \boldsymbol{q} \cdot \boldsymbol{r} - i \omega t\right]
.
\end{equation}
Here, $\boldsymbol{\xi}_0 = \left( \left| \boldsymbol{\xi}_0 \right| / \sqrt{2}, -\left| \boldsymbol{\xi}_0 \right| / \sqrt{2}, 0 \right) $ is a constant vector and $\boldsymbol{q} = \left( \left| \boldsymbol{q}_0 \right| /\sqrt{2}, \left| \boldsymbol{q}_0 \right|/\sqrt{2}, 0 \right)$ is the wavevector of the ultrasonic waves.
Using Eq.~(\ref{xi_vector}), we can show that both the symmetric strain $\varepsilon_{x^2-y^2} = \varepsilon_{xx} - \varepsilon_{yy}$ and the antisymmetric rotation $\omega_{xy}$ of the crystal lattice are induced in the crystal as follows
\cite{Kurihara_JPSJ86, Luthi_Text}:
\begin{equation}
\varepsilon_{x^2-y^2}
= \left( \frac{\partial \xi_x}{ \partial x} - \frac{\partial \xi_y}{ \partial y} \right) 
\neq 0,
\end{equation}
\begin{equation}
\omega_{xy}
= \frac{1}{2}\left( \frac{\partial \xi_y}{ \partial x} - \frac{\partial \xi_x}{ \partial y} \right) 
\neq 0
.
\end{equation}
Therefore, we should take these contributions into account for the ultrasonic attenuation coefficient $\alpha_\mathrm{T}$ and the elastic constant $C_\mathrm{T}$.
We note that both the strain and the rotation propagate through the crystal with the sound velocity $v_\mathrm{T} = \sqrt{C_\mathrm{T} / \rho}$.
However, the restoring force associated with $\omega_{xy}$ is negligible because of the rotational invariance of the elastic energy
\cite{Kurihara_JPSJ86}.

To determine which electronic degrees of freedom cause the slowing down, we focus on the group-theoretical considerations.
Because Ru-$4d$ electrons contribute to the electronic structure of Sr$_2$RuO$_4$
\cite{Bergemann_PRL84, Oguchi_PRB51},
we consider the point group symmetry $D_{4h}$ at the Ru site.
As shown in Table \ref{table_irreps}
\cite{Inui_Group, Kurihara_JPSJ86},
we can describe the bilinear coupling between $\omega_{xy}$ and the electric hexadecapole $H_z^\alpha$ both with the irrep $A_\mathrm{2g}$, as well as the strain $\varepsilon_{x^2-y^2}$ and the electric quadrupole $O_{x^2-y^2}$ both with the irrep $B_\mathrm{1g}$.
These symmetry considerations indicate that $\alpha_\mathrm{T}$ and $C_\mathrm{T}$ contain contributions from $H_z^\alpha$ and $O_{x^2-y^2}$.
Because the elastic constant and the ultrasonic attenuation coefficient satisfy the Kramers-Kronig relation, an increase in $\alpha_\mathrm{T}$ is generally accompanied by a softening of $C_\mathrm{T}$.
However, $C_\mathrm{T}$ of Sr$_2$RuO$_4$ does not exhibit additional softening at low temperatures, except below $T_\mathrm{c}$.
These observations suggest that the ultrasonic attenuation in $\alpha_\mathrm{T}$ is likely to be associated with the slowing down of hexadecapole fluctuations rather than quadrupole ordering, and is unlikely to be explained solely by the anisotropic viscosity tensor
\cite{Lupien_PRL86}.
We also note that the increase in $\alpha_\mathrm{T}$ due to enhanced electrical conductivity $\sigma$ at low temperatures
\cite{Luthi_Text}
is unlikely to account for the present results because the observed ultrasonic attenuation exhibits strong symmetry selectivity.
Further investigations of the relationship between ultrasonic attenuation and electrical conductivity would be desirable to clarify this point.

\begin{table}[b]
\begin{center}
\caption{
Irreducible representations (irreps) of deformations in the point group $D_{4h}$ for the gerade symmetry, including symmetry strains and rotations, together with their basis functions, electronic degrees of freedom (EDFs), and corresponding response functions (RFs), such as elastic constants and ultrasonic attenuation coefficients.
The angular momentum operators are defined as follows.
$l_x = -i \left( y\partial_z - z \partial_y \right)$, 
$l_y = -i \left( z \partial_x - x \partial_z \right)$, 
and 
$l_z = -i \left( x \partial_y - y \partial_x \right)$.
$C_\mathrm{B}$ and $C_u$ denote the bulk modulus associated with the isotropic volume strain $\varepsilon_\mathrm{B} = \varepsilon_{xx} + \varepsilon_{yy} + \varepsilon_{zz}$ and the elastic constant for the tetragonal strain $\varepsilon_u = 2\varepsilon_{zz} - \varepsilon_{xx} - \varepsilon_{yy}$, respectively.
}
\label{table_irreps}
\begin{tabular}{ccccc}
\hline
irrep
    &deformation
        &basis function
            &EDF
                &RF
\\
\hline
$A_\mathrm{1g}$
     &$\varepsilon_\mathrm{B}$
        &$r^2$
            &$O_\mathrm{B}$
                &$C_\mathrm{B}$
\\

     &$\varepsilon_u$
        &$3z^2 - r^2$
            &$O_{3z^2 - r^2}$
                &$C_u$
\\
$A_\mathrm{2g}$
    &$\omega_{xy}$
        &$xy \left( x^2 - y^2 \right)$
            &$H_z^\alpha$, $l_z$
                &$\alpha_\mathrm{T}$, $\alpha_{66}$
\\
$B_\mathrm{1g}$
     &$\varepsilon_{x^2-y^2}$
        &$x^2-y^2$
            &$O_{x^2 - y^2}$
                &$C_\mathrm{T}$, $\alpha_\mathrm{T}$
\\
$B_\mathrm{2g}$
    &$\varepsilon_{xy}$
        &$xy$
            &$O_{xy}$
                &$C_{66}$, $\alpha_{66}$
\\
$E_\mathrm{g}$
    &$\left\{\varepsilon_{yz}, \varepsilon_{zx} \right\}$
        &$\left\{yz, zx \right\}$
            &$\left\{ O_{yz}, O_{zx} \right\}$
                &$C_{44}$, $\alpha_{44}$
\\

    &$\left\{\omega_{yz}, \omega_{zx} \right\}$
        &
            &$\left\{ l_x, l_y \right\}$
                &$\alpha_{44}$
\\  \hline
\end{tabular}
\end{center}
\end{table}

To understand the electronic states carrying the electric hexadecapole $H_z^\alpha$ in Sr$_2$RuO$_4$, we refer to the iron-based superconductor as one of the reference systems.
In Ba(Fe,Co)$_2$As$_2$ with a superconducting transition temperature of 23 K, spin-singlet two-electron states are formed from degenerate $yz$ and $zx$ orbitals.
These states are mediated by ferro-type quadrupole–quadrupole interactions and carry the hexadecapole degree of freedom.
As a result, critical slowing down toward the superconducting transition temperature occurs due to ferro-type hexadecapole ordering.
The parameters obtained from the fitting of $\tau$ in Ba(Fe,Co)$_2$As$_2$ are summarized in Table \ref{table_parameters}
\cite{Kurihara_JPSJ86}.
In Sr$_2$RuO$_4$, the $C_\mathrm{T}$ softening above 35 K may enhance the quadrupole-quadrupole interaction.
This enhancement may promote the development of such states below approximately 15 K.
On the other hand, near the superconducting transition, no additional softening is observed.
As a result, hexadecapole ordering is not realized, although fluctuations of $H_z^\alpha$ are enhanced.

Since the slowing down of the hexadecapole fluctuations due to two-electron states can be related to the elastic softening, we discuss the origin of the softening of $C_\mathrm{T}$ in Sr$_2$RuO$_4$.
Based on the multiband electronic structure, the degenerate $yz$ and $zx$ orbitals can contribute to the elastic softening of $C_\mathrm{T}$, because the decomposition of the direct product of the irrep $E_\mathrm{g}$ contains the irrep $B_\mathrm{1g}$, as shown in Table \ref{table_Product}. 
These orbitals mainly contribute to the $\alpha$ and $\beta$ FSs \cite{Bergemann_PRL84, Oguchi_PRB51, Tsuchiizu_PRB91}.
However, several studies have proposed that the $\gamma$-FS, primarily derived from the $xy$ orbital, plays a dominant role in the response to external strain
\cite{Steppke_Science355, Hicks_Science344, Li_Nature607},
because it produces a large density of states near the M point at the Brillouin-zone boundary, a feature known as a van Hove singularity.
Therefore, the deformation of the $\gamma$-FS is likely to be crucial for the elastic softening of $C_\mathrm{T}$.
Although the contributions of $\alpha$- and $\beta$-FSs to the elasticity can be minor, the density of states of $yz$ and $zx$ orbitals can be enhanced due to the van Hove singularity through spin-orbit-induced hybridization
\cite{Tamai_PRX9}.
This behavior may also promote the development of two-electron states. 
To confirm this scenario, it is necessary to investigate the effect of external strain on the elastic softening.

Finally, we discuss possible multi-component superconducting order parameters in Sr$_2$RuO$_4$.
Our results demonstrate that the increase in $\alpha_\mathrm{T}$ is not caused by superconductivity itself, but rather by the slowing down of hexadecapole fluctuations.
Therefore, superconducting order parameters whose direct products include the irrep $A_\mathrm{2g}$ are unlikely to account for the observed ultrasonic attenuation.
Based on group-theoretical considerations summarized in Table \ref{table_Product} and elastic and superconducting studies
\cite{Ghosh_NatPhys17, Benhabib_NatPhys17},
multi-component superconducting order parameters belonging either to the irrep $E_\mathrm{g}$ or to the irreps $B_\mathrm{1g}$ and $A_\mathrm{2g}$ appear to be crucial for the elastic anomaly in $C_{66}$ with the irrep $B_\mathrm{2g}$.
This is because the decomposition of their direct product includes the irrep $B_\mathrm{2g}$, allowing bilinear coupling with the strain $\varepsilon_{xy}$.
On the other hand, combinations such as $B_\mathrm{1g}$ and $B_\mathrm{2g}$, $A_\mathrm{1g}$ and $A_\mathrm{2g}$, and $E_\mathrm{g}$ can generate the irrep $A_\mathrm{2g}$ through their direct products.
In contrast, the combination $B_\mathrm{1g}$ and $A_\mathrm{2g}$ remains compatible with the present experimental results because its direct product produces the irrep $B_\mathrm{2g}$ rather than $A_\mathrm{2g}$.

\begin{table}[htbp]
\caption{
Product table for the gerade irreducible representations of the point group $D_{4h}$
\cite{Inui_Group}
}
\label{table_Product}
\begin{tabular}{c|ccccc}
\hline
 
	&$A_\mathrm{1g}$
		&$A_\mathrm{2g}$
			&$B_\mathrm{1g}$
				&$B_\mathrm{2g}$
                    &$E_\mathrm{g}$
\\
\hline
$A_\mathrm{1g}$
	&$A_\mathrm{1g}$
		&$A_\mathrm{2g}$
			&$B_\mathrm{1g}$
				&$B_\mathrm{2g}$
                    &$E_\mathrm{g}$
\\
$A_\mathrm{2g}$ 
	&$A_\mathrm{2g}$
		&$A_\mathrm{1g}$
			&$B_\mathrm{2g}$
				&$B_\mathrm{1g}$
                    &$E_\mathrm{g}$
\\
$B_\mathrm{1g}$ 
	&$B_\mathrm{1g}$
		&$B_\mathrm{2g}$
			&$A_\mathrm{1g}$
				&$A_\mathrm{2g}$
                    &$E_\mathrm{g}$
\\
$B_\mathrm{2g}$ 
	&$B_\mathrm{2g}$
		&$B_\mathrm{1g}$
			&$A_\mathrm{2g}$
				&$A_\mathrm{1g}$
                    &$E_\mathrm{g}$
\\
$E_\mathrm{g}$ 
	&$E_\mathrm{g}$
		&$E_\mathrm{g}$
			&$E_\mathrm{g}$
				&$E_\mathrm{g}$
                    &$A_\mathrm{1g} \oplus A_\mathrm{2g} \oplus B_\mathrm{1g} \oplus B_\mathrm{2g}$
\\
\hline
\end{tabular}
\end{table}

Our ultrasonic attenuation measurements provide new insights into the elasticity and superconductivity of Sr$_2$RuO$_4$.
Further comprehensive studies of low-temperature dynamical properties, as well as theoretical calculations clarifying the slowing down of $H_z^\alpha$ fluctuations associated with the formation of two-electron states, would be highly desirable.
On the other hand, theoretical studies have proposed the possibility of $A_\mathrm{2g}$-type hexadecapole ordering in localized $5f^2$ electrons systems
\cite{Kusunose_JPSJ80}. 
To deepen the understanding of the slowing down of hexadecapole fluctuations in Sr$_2$RuO$_4$, further theoretical studies considering its multiband electronic structure and possible multipole degrees of freedom would also be desirable.


\section{
Conclusion and perspective
}
\label{conclusion}

In this study, we performed ultrasonic measurements on the unconventional superconductor Sr$_2$RuO$_4$.
We found that the ultrasonic attenuation coefficient $\alpha_\mathrm{T}$ for the in-plane transverse ultrasonic waves associated with the elastic constant $C_\mathrm{T} = \left( C_{11} - C_{12} \right)/2$ exhibits an anomalous increase at low temperatures, while the elastic softening of $C_\mathrm{T}$ expected from the Kramers-Kronig relation is absent.
Based on the Landau-Khalatnikov theory, we found that fluctuations of the electronic degrees of freedom show slowing down, reaching a relaxation time of $\tau \sim 0.14 \times 10^{-9}$ s ($\tau^{-1} \sim 7.1$ GHz) at the superconducting transition temperature of $T_\mathrm{c} = 1.42$ K.
This slowing down follows the Ising universality with a critical exponent product $z \nu = 1$ as expected from mean-field theory.
Group-theoretical considerations suggest that the electric hexadecapole $H_z^\alpha$ with the irrep $A_\mathrm{2g}$ in the point group $D_{4h}$ is the electronic degree of freedom responsible for the slowing down.
Our ultrasonic measurements also demonstrated that the rotation $\omega_{xy}$ introduced by the in-plane transverse waves plays an essential role in Sr$_2$RuO$_4$ in addition to the strains.
These results establish ultrasonic attenuation measurements as a powerful probe of dynamical electronic multipole fluctuations and provide new constraints on the symmetry and microscopic nature of superconductivity in Sr$_2$RuO$_4$.

\section*{Acknowledgments}
Part of this work is based on the Master's thesis of Shunsuke Yaoita, submitted in partial fulfillment of the requirements for the Master of Science degree at the Graduate School of Science and Technology, Niigata University, Niigata (2014).
This work was partly supported by JSPS Grants-in-Aid for Transformative Research Areas (A) (JP 24H01629) and Early-Career Scientists (JP 20K14404, JP 22K13999).

We thank Masayoshi Nozawa, Kazuya Hoshino, Satoshi Hakamada, Keigo Naito, and Haruhi Okumura for their experimental assistance.
We also thank Rie Kurihara for valuable advice on the schematic illustrating of the experimental setup.


\bibliographystyle{jpsj}
\bibliography{main}

\end{document}